\documentclass[apj]{emulateapj}
\usepackage{apjfonts}
\usepackage{graphicx}
\newcommand{\thch}{$^{13}$CH}
\newcommand{\twch}{$^{12}$CH}
\newcommand\teff{T$_{\rm eff}$}
\newcommand\logg{log $g$}

\newcommand{\mteff}{\rm T_{eff}}
\newcommand{\mlogg}{\rm{log}g}
\newcommand{\solmass}{M$_{\odot}$}
\newcommand{\mem}[1]{\ensuremath{\mathrm{ #1}}}

\newcommand{\msun}{\ensuremath{\, {\rm M}_\odot}}

\shorttitle{Nitrogen in C-enhanced stars}
\shortauthors{Johnson et al.}

\begin{document}
\title{A Search for Nitrogen-Enhanced Metal-Poor Stars\altaffilmark{1}}

\author{Jennifer A.\ Johnson\altaffilmark{2}}
\affil{Dominion Astrophysical Observatory, Herzberg Institute of Astrophysics,
  National Research Council, 5071 West Saanich Rd., Victoria, BC V9E 2E7,
  Canada} 
\email{Jennifer.Johnson@nrc-cnrc.gc.ca}

\author{Falk Herwig}
\affil{Theoretical Astrophysics Group, Los Alamos National Laboratory, Los
  Alamos, NM 87545, USA} 
\email{fherwig@lanl.gov}

\author{Timothy C. Beers}
\affil{Department of Physics \& Astronomy, CSCE: Center for the Study of
  Cosmic Evolution, and JINA: Joint Institute for Nuclear Astrophysics,
  Michigan State University, East Lansing, MI 48824, USA}
\email{beers@pa.msu.edu}
\and

\author{Norbert Christlieb}
\affil{Hamburger Sternwarte, Universit\"at Hamburg, Gojenbergsweg 112, D-21029
  Hamburg, Germany}
\email{nchristlieb@hs.uni-hamburg.de}

\altaffiltext{1}{Based on observations obtained at Cerro
Tololo Inter-American Observatory and Kitt Peak National Observatory, a 
division of the National Optical Astronomy Observatories, which
are operated by the Association of Universities for Research
in Astronomy, Inc. under cooperative agreement with the National Science
Foundation.}
\altaffiltext{2}{Present address: Department of Astronomy, Ohio State
University, 140 West 18th Avenue, Columbus, OH 43210}

\begin{abstract}
  Theoretical models of very metal-poor intermediate-mass Asymptotic Giant
  Branch (AGB) stars predict a large overabundance of primary nitrogen. The
  very metal-poor, carbon-enhanced, s-process-rich stars, which are thought to
  be the polluted companions of now-extinct AGB stars, provide direct tests of
  the predictions of these models. Recent studies of the carbon and nitrogen
  abundances in metal-poor stars have focused on the most carbon-rich stars,
  leading to a potential selection bias against stars that have been polluted
  by AGB stars that produced large amounts of nitrogen, and hence have small
  [C/N] ratios. We call these stars Nitrogen-Enhanced Metal-Poor (NEMP) stars, and 
define them as having [N/Fe] $> +0.5$ and [C/N] $< -0.5$. 
In this paper, we report on the [C/N] abundances of a sample
  of 21 carbon-enhanced stars, all but three of which have ${\rm [C/Fe]} <
  +2.0$. If NEMP stars were made as easily as
  Carbon-Enhanced Metal-Poor (CEMP) stars, then we expected to find between two and
  seven NEMP stars. Instead, we found no NEMP stars in our sample. Therefore,
  this observational bias is not an important contributor to the apparent
  dearth of N-rich stars. Our [C/N] values are in the same range as values
  reported previously in the literature ($-0.5$ to $+2.0$), and all stars 
are in 
disagreement with the predicted
  [C/N] ratios for both low-mass and high-mass AGB stars. We suggest that
  the decrease in [C/N] from the low-mass AGB models is due to enhanced
  extra-mixing, while the lack of NEMP stars may be caused by unfavorable mass
  ratios in binaries or the difficulty of mass transfer in binary systems
  with large mass ratios.

\end{abstract}

\section{Introduction}\label{Sect:Intro}

Very Metal-Poor (VMP) stars ($\mem{[Fe/H]}\le -2.0$)\footnote{We adopt the usual
spectroscopic notation that [A/B] $\equiv$ {\rm log}$_{\rm 10}$(N$_{\rm
A}$/N$_{\rm B}$)$_{\star}$ -- {\rm log}$_{\rm 10}$(N$_{\rm A}$/N$_{\rm B}$)
$_{\odot}$, and log~$\epsilon$(A) $\equiv$ {\rm log}$_{\rm 10}$(N$_{\rm
A}$/N$_{\rm H}$) + 12.0, for elements A and B.} provide essential tools for the
study of element production in stars and galactic chemical processing during the
early stages of the evolution of our Galaxy. The number of VMP stars with
measured elemental abundance ratios has been increasing rapidly in the past
decade. The large modern surveys for metal-poor stars, most importantly the HK
survey of Beers and colleagues \citep{beers:92,beers:99b} and the Hamburg/ESO
Survey (HES) of Christlieb and collaborators \citep{christlieb:03}, have
produced medium-resolution confirmation spectra of many thousands of metal-poor
candidates; the most interesting of these have been (and are being) followed up
with high-resolution spectroscopic studies on large-aperture telescopes
\citep[e.g.  ][]{cayrel:04,honda:04b,johnson:04,barklem:05}.
  
One of the most important discoveries of these new surveys is that 
at least 20\,\% of all VMP stars exhibit conspicuous enrichments of the CNO
elements, most notably C \citep{beers:05,lucatello:06}.
These Carbon-Enhanced Metal-Poor (CEMP) stars are defined by Beers \& Christlieb
as metal-poor stars with $\mem{[C/Fe]} > +1.0$. Many of the CEMP stars
exhibit overabundances of the elements associated with s-process
nucleosynthesis. Aoki et al. (2003) estimate this fraction to be between 70\,\%
and 80\,\%; Beers \& Christlieb refer to these as the CEMP-s stars. The
observed abundance patterns for CEMP-s stars suggest nucleosynthetic origin in low- or
intermediate-mass stars that have evolved through the thermally pulsing
asymptotic giant branch (AGB) phase, and later transferred this processed
material to a surviving low(er)-mass companion. In this sense the CEMP-s stars
are the metal-poor analogs of the classical CH stars \citep{keenan:42}. The AGB
star that originally enriched the presently observed companion VMP star is now a
white dwarf; its presence is often revealed by tell-tale radial-velocity
variations observed for the companion star. Based on the still-limited numbers
of CEMP-s stars for which sufficient multi-epoch spectroscopic data has been
obtained, a $100$\,\% binary fraction is possible \citep{lucatello:05}. It is
thus expected that the large overabundances of a number of elements observed in
CEMP-s stars should reflect the AGB nucleosynthetic yields, perhaps with some
modification due to giant-branch evolution of the companion.

In all AGB stars C is produced by the triple-$\alpha$ reaction in He-shell
flashes and convectively dredged up into the stellar envelope. For stars of
low initial mass (2--3\,\msun), this eventually leads to a C-rich composition,
with ${\rm C/O} > 1$. In these models N is not enhanced because it is burned
during the He-shell flash. During the interpulse phase the convective envelope
has no mixing connection with the H-shell, according to standard models, and
no alteration of N is expected. For larger masses ($>3.5\,\msun$, depending on
metallicity) the efficiently dredged up carbon is transformed into N by the
hot-bottom burning (HBB) process
\citep{lattanzio:92b,boothroyd:93,forestini:97,herwig:04a,ventura:05}. As a
result, the models predict small C/N ratios in the stellar envelope
($\mem{[C/N]} \approx -1$).

\begin{figure}
\includegraphics[width=8cm]{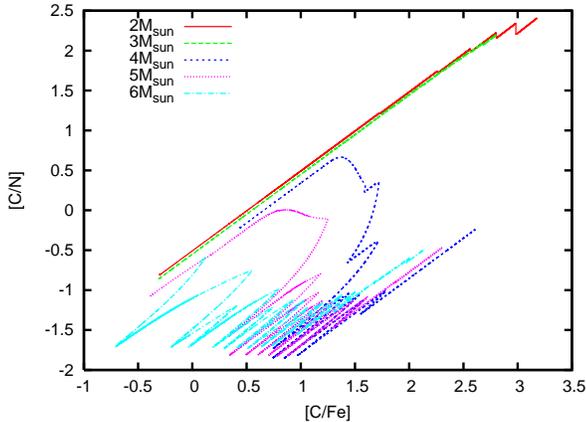}
\caption{\label{Fig:AGBabund}Surface
  abundances of AGB stars of different masses for metallicity [Fe/H] = $-2.3$
 over several dredge-up episodes.
  As more dredge-up occurs, the surface abundance of C increases. For the
  higher-mass cases, dredged-up C is processed into N via HBB, leading to lower
  C/N ratios.  The final total yields of the elements produced by the AGB
  stars and ejected into the ISM by mass loss are given in \citet{herwig:04b}.}
\end{figure}

Simulations of low- and intermediate-mass stellar evolution confirm that the
well-established trends of C and N production in solar-metallicity AGB stars
extend to very low metallicity \citep{herwig:04b}. The amount of N produced is
independent of the initial metallicity of the star, because it is based on the
primary production of C in the He-burning shell. Therefore, HBB in
intermediate-mass VMP AGB stars provides a primary source of nitrogen in the
early Universe. Figure~\ref{Fig:AGBabund} shows the evolution of [C/N] ratios
on the surface of metal-poor AGB stars from the models of \citet{herwig:04b}.
Two effects are immediately clear: (1) a lower predicted [C/N] ratio in more
massive AGB stars, and (2) a lower predicted [C/Fe] on the surfaces of these
same stars.


\begin{figure}
\includegraphics[width=8cm]{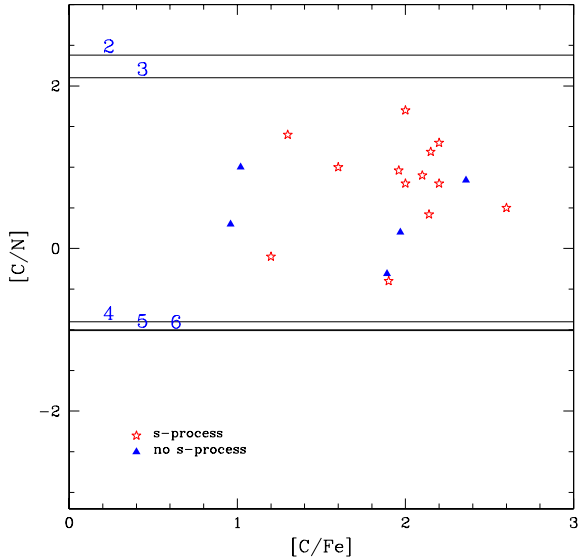}
\caption{\label{Fig:CFeCNold}
  Literature values for [C/N] ratios in VMP as of 2002. The
  lines are labeled with the masses of the AGB stars from the models of
  \citet{herwig:04b}. We assume that the initial C and N on the observed
  star's surface are overwhelmed by the contribution from the AGB star.
  However, because the [C/Fe] ratio depends on both the amount of carbon
  transferred from the AGB star and the amount of Fe on the companion star
  surface, there is no {\it a priori} dependence of the [C/Fe] ratio on the
  mass of the AGB star. We do not include in this figure 
CS~22892-052, which is r-process rich, CS~22949-037, which is
extremely O-rich, or CS~30314-067, which has new abundances reported
in this paper.}

\end{figure}

The predicted time-averaged yields for AGB stars are shown in
Figure~\ref{Fig:CFeCNold}. In this Figure we also plot [C/N] ratios for CEMP 
stars that were available in the literature when we began this project. We
would expect that at least one-third of the stars should have been polluted by
intermediate-mass AGB stars, assuming (1) stars between 3.5\,{\solmass} and
7.5\,{\solmass} undergo HBB, (2) stars between 1.0\,{\solmass} and
7.5\,{\solmass} produce enough C to enhance their companions
during mass transfer, and (3) the binary mass ratios from \citet{pinn:06} are
correct for low-metallicity stars. It is evident that the observed [C/N]
ratios in CEMP stars do not fit the predictions of either the low-mass or the
intermediate-mass AGB models, but instead fall in an intermediate regime that
is not covered by the models.

We are left with two questions. First, why do the observations of stars with large [C/Fe]
show larger N (smaller [C/N]) than predicted by models for the evolution of
2--3\,\msun\ stars? This problem is reminiscent of the high N abundances
observed in the moderately metal-poor CH stars, and may be related to mixing
processes that are not accounted for in models of AGB evolution
\citep{vanture:92b}. This issue will be discussed later in this paper. Second,
where are the VMP stars that were, in the context of the mass-transfer scenario,
polluted by the N-rich 3.5--7.5\,\msun\ stars? The solution for this problem is
less obvious. The HBB responsible for the very efficient production of N in
intermediate-mass stars is a robust prediction of stellar-evolution models. If
the binary mass-transfer scenario is invoked for the CEMP-s stars with ${\rm
[C/N]} \approx 1$, then one may ask why there are no NEMP
stars with ${\rm [C/N]} \approx -1$, as might be expected to arise from
systems in which the donor is a more massive HBB AGB star. Before
considering possible solutions to this second problem we need to test one
obvious possibility, that the absence of NEMP stars among previously analysed
VMP stars may be simply the result of a selection bias against their detection.

Because of the requirement for large-aperture telescopes in order to obtain high-resolution,
high-$S/N$ spectra of the CEMP stars in the HK and HES, only a limited number
of these objects have been observed to date. Special attention has been
focused on the subsample of CEMP stars that are likely to be the most
iron-poor and/or carbon-rich \citep[e.g.][]{norris:97b,aoki:02b}. This may
have led to an observational bias against the discovery of NEMP stars. The CH
features at 4305\,{\AA} are routinely covered in the medium-resolution
spectroscopic follow-up of metal-poor candidates from the HK survey and the
HES; thus, CEMP stars are readily identified and placed on target lists for
examination at high resolution. However, N-rich stars are not so easily
recognized. The medium-resolution confirmation spectra generally extend no
bluer than about 3600\,{\AA}; the only N-sensitive feature included in this
range is the CN band at 3850\,{\AA}, which requires large enhancements of {\it
  both} C and N to be strong.  As a result, the majority of stars in the
current literature with detailed studies of their elemental abundance patterns
are those with ${\rm [C/N]} > 0$, i.e., they are more C-rich than they are
N-rich. Thus, the lack of stars with ${\rm [C/N]} < 0$ could well be an
observational selection effect, especially since the lower C-enhancements
predicted in intermediate-mass AGB stars would make their companions
potentially less C-rich, and hence less likely to be followed up at high spectral-resolution.

Previous low-resolution surveys of metal-poor (${\rm [Fe/H]} <-1.0$) dwarf
stars that included the blue NH band have discovered some stars with stronger
than average nitrogen abundances, including HD~74000 and HD~160617
\citep{bessell:82}, HD~97916 and HD~166913 \citep{laird:85} and HD~25329
\citep{carbon:87}. Higher-resolution follow-up studies
\citep[e.g.,][]{beveridge:94,mashonkina:03} have shown that these stars are
somewhat s-process rich (${\rm [Ba/Fe]}\sim +0.5$). They do not appear to be
C-enhanced, however, and their binary status is not confirmed. These stars are
probably related to the phenomenon discussed here, but we confine ourselves in
the rest of the paper to the discussion of CEMP stars
found in recent surveys.

We have undertaken a medium-resolution observing campaign to address the
question of whether observational selection biases might be responsible for
the lack of known NEMP stars. We examine a sample of VMP stars with moderate
carbon enhancements, $+0.5 \le \rm{[C/Fe]} \le +1.0$ (based on their
medium-resolution confirmation spectra), in order to better constrain the
range of the N enhancements in these stars, and to see if the N abundances
obtained agree better with expectations based on AGB models. This is
accomplished using near-UV medium-resolution spectroscopy that covers the
region of the NH band at 3360--3370\,{\AA} for a sample of 21 moderately
carbon-enhanced metal-poor stars, along with a number of similar stars from
the literature with available high-resolution results. In
\S\,\ref{Sect:Observations} we describe our sample selection criteria,
observations, and data reduction procedures. Details of our abundance analysis
for this sample are provided in \S\,\ref{Sect:AbundanceAnalysis}. In
\S\,\ref{Sect:Results} we summarize our results. A discussion of the
theoretical expectations and a comparison with our present results is provided
in \S\ref{Sect:Discussion}.

\section{Observations and Data Reduction}\label{Sect:Observations}

\subsection{Sample Selection}

We wished to observe a sample of stars with ${\rm [Fe/H]} <-2.5$ and [C/Fe]
between $+0.5$ and $+1.0$. Our primary source of targets was the list of
[Fe/H] and [C/Fe] values for the HK survey stars calibrated by
\citet{rossi:05}. We also added stars from the Beers, Preston, \& Shectman
(1992) list of stars with abnormally strong G-bands (their Table 8). To
compare our derived abundances with those obtained from higher-resolution
spectra, we also took spectra of some well-known bright metal-poor HD and BD
stars, as well as three stars from the HK survey, namely
CS~22892-052, CS~22968-014, and CS~22950-46. These three are not
part of our C-enhanced sample either, because they are known to be
r-process-rich and probably polluted by a different mechanism altogether
(CS~22892-052) or are not C-enhanced (CS~22968-014 and CS~22950-048).
Ideally, all of
our stars would be subgiant or main-sequence stars, to avoid possible
alterations in the surface C and N abundances due to CN processing and mixing
on the red giant branch, which can result in decreasing C and increasing N
abundances \citep{gratton:00,spite:05}. However, molecular features such as CH
and NH are much weaker in hotter stars, so we decided to include giants in our
list. Any CN cycling that might have operated would serve to increase the N
abundances, and therefore bias the study toward finding more NEMP stars than
would otherwise be the case. This turned out not to be a concern in the final
analysis.

\subsection{Observations}

\begin{figure}
\includegraphics[width=8cm]{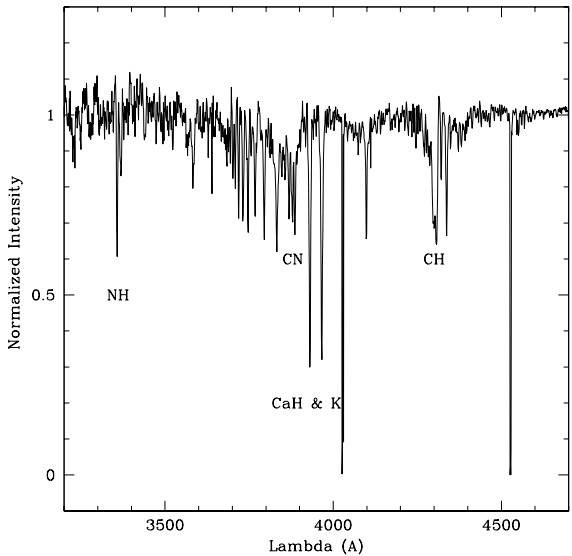}
\caption{\label{Fig:Example_Spectrum} A normalized spectrum of CS~22947-187
  from the RCSpec at CTIO.}
\end{figure}

To observe the NH band we require near-UV sensitivity, however, a resolution
of $2$--$2.5$\,{\AA} is more than adequate. The RC spectrographs at Kitt Peak
National Observatory and at Cerro Tololo Interamerican Observatory have these
capabilities, allowing us to observe both northern and southern hemisphere
targets. We observed 8 stars at KPNO over three nights from 30 Aug to 1 Sep
2003. We employed the BL420 grating and the CuS0$_4$ order blocking filter.
This produced a wavelength coverage of 3250\,{\AA} to 4700\,{\AA} with a
resolution of 2.0{\AA}. The F3B chip was used, with a gain setting of
2.3\,e$^{-}$/ADU, and a read noise of 7.5\,e$^-$. The dispersion was 0.76\AA/pix. A quartz lamp with a BG3
blocking filter was used to obtain flatfields, and the FeAr lamp was used for
wavelength calibration. We also observed 20 stars over 6 nights at CTIO in two
observing sessions: 6--8 Sep 2003 and 26--28 Nov 2004. Some part of each run
was lost to weather. We used the RC Spectrograph with the blue collimator, the
KPGL1 grating, and the CuSO$_4$ order-blocking filter. We adopted Decker 2,
which provided a 1 arcsec slit. This produced a wavelength coverage of
3250\,{\AA} to 4700\,{\AA} with a resolution of 2.5\,{\AA}.  The gain setting
was 1.94\,e$^-$/ADU, and the read noise was 7.5\,e$^-$. The dispersion was 0.95\AA/pix. A helium-neon-argon
lamp was used for wavelength calibration for these data. The second run at
CTIO (26--28 Nov 2004) was plagued by large amounts of scattered light in the
spectra. Although the scattered light was subtracted, the Poisson noise from
its presence reduced the $S/N$ around the NH band to $\sim 10$--$20$. We were
still able to obtain useful upper limits on N, and therefore lower limits on
[C/N], with these data. Table~\ref{Tab:ObsLog} present a summary of the
observations. Figure~\ref{Fig:Example_Spectrum} shows a typical spectrum
obtained from these observations.

\begin{deluxetable*}{lcrrrc}
\tabletypesize{\footnotesize}
\tablenum{1}\label{Tab:ObsLog}
\tablewidth{0pt}
\tablecaption{Log of Observations}
\tablehead{
\colhead{Star} & \colhead {V} & \colhead{Telescope} & \colhead{Date of
Obs.} & \colhead{Exposures} & \colhead {$S/N$ at $3400$\,{\AA}}  
}
\startdata
BD$-18^{\circ}\,5550$ & 9.270 & CTIO & 2003 Sep 05 & 2$\times$600 & 192 \\
HD~122563    & 6.196 &KPNO  & 2003 Aug 31 & 3$\times$200s & 267 \\
HD~160617    & 8.740& CTIO  & 2003 Sep 04 & 3$\times$600s & 207\\
HD~186478    & 8.920& CTIO  & 2003 Sep 04 & 3$\times$600s & 425\\
CS~22174-007 &12.409 & CTIO & 2003 Aug 30 & 3$\times$1800s & 142 \\
CS~22183-031 &13.622 & CTIO & 2004 Nov 28 &  3$\times$1200s & 17\\
CS~22879-029 &14.425 & CTIO & 2003 Sep 06 & 1$\times$1200s,2$\times$900s & 117 \\
CS~22884-097 &14.868 & CTIO & 2003 Sep 05 & 3$\times$1200s & 33 \\
CS~22887-048 &12.866 & KPNO & 2003 Aug 30 & 1$\times$1800s,2$\times$900s & 123 \\
CS~22891-171 &14.293 & CTIO & 2003 Sep 06 & 3$\times$1200s & 58 \\
CS~22892-052 &13.213 & CTIO & 2003 Sep 04 & 4$\times$600s & 115 \\
CS~22898-062 &13.788 & KPNO & 2003 Aug 31 & 3$\times$1800s & 71 \\
CS~22945-024 &14.360 & CTIO & 2003 Sep 04 & 3$\times$1200s & 55\\
CS~22947-187 &12.962 & CTIO & 2003 Sep 04 & 3$\times$1200s  & 117\\
CS~22948-104 &13.929 & CTIO & 2004 Nov 28 & 3$\times$1200s & 17\\
CS~22949-008 &14.168 & KPNO & 2003 Aug 31 & 1$\times$600s 3$\times$1800s & 95 \\
CS~22950-046 &14.224 & KPNO & 2003 Aug 30 & 4$\times$ 1800s & 45 \\
CS~22958-042 &14.516 & CTIO & 2003 Sep 06 & 4$\times$1200s & 42\\
CS~22958-083 &14.423 & CTIO & 2044 Nov 27 & 3$\times$1200 &17 \\
CS~22960-053 & 14.830& CTIO & 2003 Sep 06 & 3$\times$1200s & 40 \\
CS~22968-014 & 13.684& CTIO & 2003 Sep 04 & 3$\times$1200s & 80\\
CS~29493-090 & 14.039& CTIO & 2003 Sep 04 & 3$\times$1200s &  57 \\
CS~29495-042 & 14.516& CTIO & 2003 Sep 06 & 1$\times$1200s 2$\times$900s &67 \\
CS~29497-030 & 12.656& CTIO & 2003 Sep 06 & 1$\times$1200s,3$\times$600s & 134\\
CS~29512-073 & 14.137& KPNO & 2003 Aug 30 & 2$\times$1800s,1$\times$900 & 66 \\
CS~30314-067 & 11.817& CTIO & 2003 Sep 04 & 3$\times$1200s & 122 \\
CS~31062-041 & 13.934& KPNO & 2003 Aug 31 & 3$\times$1800s & 53 \\
CS~31080-095 & 12.989& CTIO & 2004 Nov 27 & 4$\times$1200s & 17\\
\enddata
\end{deluxetable*}

\subsection{Data Reduction}

The data were reduced using IRAF\footnote{IRAF is distributed by the National
  Optical Astronomy Observatories, which are operated by the Association of
  Universities for Research in Astronomy, Inc., under cooperative agreement
  with the National Science Foundation.}. The usual reduction procedures were
applied: bias-subtraction, flat-fielding, spectral extraction, sky
subtraction, and wavelength calibration. For most stars, with the
exception of the observations in 2004, we achieved our
desired $S/N$ of 50/1 per resolution element. The $S/N$ at the
NH bandhead is listed in Table~\ref{Tab:ObsLog}.

\section{Abundance Analysis}\label{Sect:AbundanceAnalysis}

We used the 2002 version of MOOG \citep{sneden:73} for our synthesis of the NH,
CH and \ion{Ca}{2}~K regions. We interpolated the grids of \citet{kurucz:atm},
using the models with overshooting, but without $\alpha$-enhancements or the new
opacity distribution functions, because that grid was the most extensive at the
lowest metallicities. For the accuracy we require in [C/N] at these low
metallicities, our choice of model atmosphere grid is not a significant source
of error.

\subsection{Molecular and Atomic Data}\label{Sect:AtomicData}

We assembled linelists covering 3340--3400\,{\AA} for NH and 4270-4330\,{\AA}
for CH. The atomic parameters are from the Vienna Line Database
\citep{piskunov:99}. The molecular line lists were in general adopted from
\citet{kurucz:06} and are discussed further below. We adopted the solar values
of \citet{anders:89}, with the exception of Fe, where we assume a solar value
of $\log \epsilon = 7.52$. While the solar values of C and N have been the
subject of much dispute in recent years \citep{asplund:05b}, in our case
they merely represent scaling factors applied to both the yields from AGB
stars of \citet{herwig:04b}, and to our derived $\log \epsilon$ for C and N from
these stars.

\subsubsection{NH}

We used the bandhead for the A-X (0-0) and (1-1) transition at 3360\,{\AA} and
3370\,{\AA}, respectively. The only important isotope is $^{14}$N, because the
$^{14}$N/$^{15}$N ratio is $> 100$ wherever it has been measured
\citep{chin:99}. The wavelengths of NH from \citet{kurucz:06} are in excellent
agreement with the laboratory wavelengths measured by \citet{brazier:86} for
the regions near the NH bandhead. However, \citet{shavrina:96} pointed out
that the Kurucz gf-values were too high by a factor of two. Therefore, we have
adopted Kurucz wavelengths, but divide his oscillator strengths by this
factor. These gf-values are then on the same scale as those of
\citet{sneden:73}.

Conflicting values for the dissociation potential of NH exist in the
literature. \citet{seal:66} measured $3.21 \pm 0.16$\,eV. More recent
experiments have found higher values. \citet{graham:78} measured $D_0 \leq
3.47 \pm 0.05$\,eV, while \citet{tarroni:97} measured $D_0 \geq 3.419 \pm
0.010$\,eV. With these results in mind, we have adopted 3.45\,eV, close to
the value used by \citet{spite:05} in their study of C, N and O abundances in VMP stars.

\begin{figure}
\includegraphics[angle=270,width=8cm]{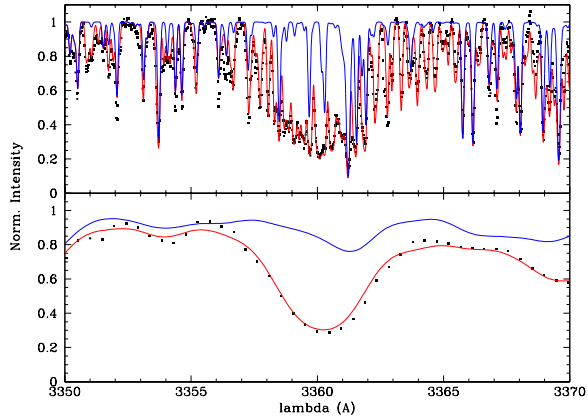}
\caption{\label{Fig:NHsynth} Synthesis of the NH bandhead region in HD~186478 for
  a high-resolution spectrum (top) and a low-resolution spectrum (bottom). The
  model atmosphere used in this paper was adopted for both syntheses. One line
  shows the synthesis without any N, while the second shows the synthesis with
  log$\epsilon$(N)=6.12.  Filled squares show the data.}
\end{figure}

Our linelist in the NH region is illustrated in Figure~\ref{Fig:NHsynth}. We
first show the fit to a high-resolution spectrum of the metal-poor star
HD~186478 from \citet{johnson:02a} and then to the lower-resolution spectrum
taken for this study.

If we used a dissociation potential of $3.21$\,eV instead of $3.45$\,eV, our
derived N abundances would increase by $\sim 0.3$\,dex. If we used the
uncorrected Kurucz gf-values, our abundances would decrease by a similar amount.
The lack of agreement among different studies on the dissociation potential
and oscillator strengths is a major source of the differences in the derived C
and N abundances in the present literature. Fortunately, these issues are
dwarfed by the $> 1$\,dex difference we hope to observe.

\subsubsection{CH}

\begin{figure}
\includegraphics[angle=270,width=8cm]{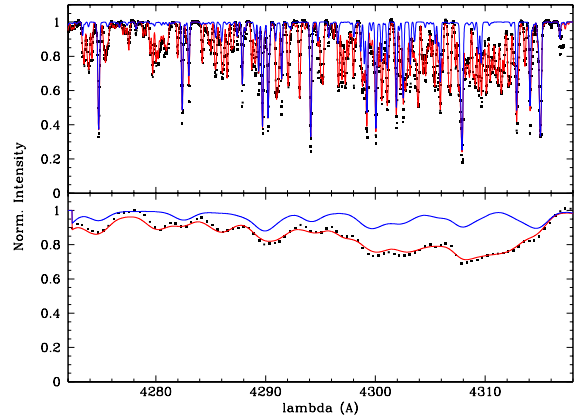}
\caption{\label{Fig:CHsynth} Synthesis of the CH bandhead region in HD~186478 for
  a high-resolution spectrum (top) and a low-resolution spectrum (bottom). The
  model atmosphere used in this paper was adopted for both syntheses. One line
  shows the synthesis without any C, while the second shows the synthesis with
  $\log\epsilon ({\rm C})=5.73$.  Filled squares show the data.}
\end{figure}

Our analysis of the carbon abundances for program stars used the CH G band
feature. Again, we used the Kurucz line list for this region. The wavelengths
and oscillator strengths agreed well with the more limited lists of
wavelengths from \citet{zach:95,zach:97} and oscillator strengths from LIFBASE
\citep{luque:96}. Theoretical models predict different $^{12}$C/$^{13}$ ratios
for AGB stars of different masses, ranging from $> 1000$ for 2\,{\solmass} to
$\sim 5$ for 6\,{\solmass}.  The G band contains both \twch{} and \thch{}
lines.  However in our low-resolution spectra, these lines are intermingled to
such an extent that if the total carbon abundance does not change, our
synthesis of the G-band remains essentially independent of the
carbon isotope ratios. We adopted a value of 80 for our synthesis.
Figure~\ref{Fig:CHsynth} shows the fit in the G-band region for HD~186478.

\subsubsection{C and N Abundances from Different Molecular Species}

Most studies of metal-poor stars measure the N abundance from the blue CN
system near 3870--3880\,{\AA}, as this wavelength region is usually covered in
the setups used for medium-resolution spectroscopy. \citet{spite:05} were able
to work with high-resolution spectra of metal-poor stars that covered both the
NH and CN features. They found that the NH lines consistently yielded $\sim
0.3$\,dex higher N abundances than those obtained from the CN lines. The
source of the disagreement is unclear (and we do not resolve it here), but
note that our N abundances (derived from NH) would be lower by 0.3\,dex if the
CN abundance scale were the correct one. Literature sources also derive C
abundances from both the C$_2$ and the CH features. \citet{aoki:02c} measured
a higher C abundance, by 0.2\,dex, in the subgiant LP~625-44 from the C$_2$
lines as compared to the CH feature. These differences could result in
systematic offsets in C or N abundances; however, their magnitude is smaller
than the size of the expected [C/N] range. Where necessary for our discussion,
we thus adopt literature C and N values as stated, selecting C from CH and N
from NH whenever possible.

\subsection{Model Atmosphere Parameters}

We used broadband photometry, provided by either \citet{beers:06a} or taken
from the SIMBAD listing, to derive \teff\ and isochrones to
derive gravities. We also took advantage of the well-known correlation between {\logg}
and $\xi$ for estimation of microturbulent velocities, as described below. Finally,
we synthesized the \ion{Ca}{2} K line to determine the appropriate metallicity.
We discuss additional details and comparisons with other methods
of deriving model atmosphere parameters below.

\begin{deluxetable*}{lrrrrrrrr}
\tabletypesize{\footnotesize}
\tablenum{2}\label{Tab:Photometry}
\tablewidth{0pt}
\tablecaption{Photometry Data}
\tablehead{
\colhead{Star} & \colhead {V} & \colhead{B-V} & \colhead{U-B}
 & \colhead{V-R} & \colhead {V-I} & \colhead{V-J} & \colhead{V-K} 
& \colhead{E(B-V)}   
}
\startdata
BD$-18^{\circ}\,5550$   &  9.270 &   9.270 &   0.910 &   0.230 &   0.530 &   2.068 &   2.715 &  0.17 \\
HD~122653    &  6.196 &   0.912 &   0.340 &   0.580 &   \dots &   1.815 &   2.473 &  0.02 \\
HD~160617    &  8.740 &   \dots &   \dots &   \dots &   \dots &   1.112 &   1.429 &  0.01 \\
HD~186478    &  8.920 &   0.930 &   \dots &   \dots &   \dots &   1.803 &   2.477 &  0.10 \\
CS~22174-007 & 12.409 &   0.694 &   \dots &   0.448 &   0.900 &   1.524 &   2.091 &  0.03 \\
CS~22183-031 & 13.622 &   0.673 &   \dots &   0.444 &   0.898 &   1.511 &   2.040 &  0.04 \\
CS~22884-097 & 14.868 &   0.579 &   \dots &   0.407 &   0.849 &   1.401 &   1.742 &  0.23 \\
CS~22887-048 & 12.866 &   0.389 &   \dots &   0.278 &   0.556 &   0.901 &   1.192 &  0.05 \\
CS~22879-029 & 14.425 &   0.420 &   \dots &   0.280 &   0.604 &   1.000 &   1.299 &  0.04 \\
CS~22891-171 & 14.293 &   0.872 &   \dots &   0.485 &   0.918 &   1.522 &   2.030 &  0.07 \\
CS~22892-052 & 13.213 &   0.800 &   \dots &   0.491 &   0.990 &   1.721 &   2.284 &  0.03 \\
CS~22898-062 & 13.788 &   0.627 &   \dots &   0.420 &   0.840 &   1.439 &   1.937 &  0.05 \\
CS~22945-024 & 14.360 &   0.722 &$-$0.006 &   \dots &   \dots &   1.450 &   1.926 &  0.03 \\
CS~22947-187 & 12.962 &   0.648 &   \dots &   0.454 &   0.877 &   1.494 &   1.934 &  0.09 \\
CS~22948-104 & 13.929 &   0.604 &   \dots &   0.410 &   0.821 &   1.431 &   1.917 &  0.02 \\
CS~22949-008 & 14.168 &   0.494 &   \dots &   0.310 &   0.623 &   1.045 &   1.380 &  0.04 \\
CS~22950-046 & 14.224 &   \dots &   \dots &   0.585 &   1.158 &   1.974 &   2.641 &  0.06 \\
CS~22958-042 & 14.516 &   0.479 &   \dots &   0.294 &   0.614 &   1.001 &   1.303 &  0.02 \\
CS~22958-083 & 14.423 &   0.664 &$-$0.030 &   0.442 &   0.902 &   1.477 &   2.029 &  0.04 \\
CS~22960-053 & 14.830 &   0.760 &   0.100 &   \dots &   \dots &   1.532 &   2.069 &  0.01 \\
CS~22968-014 & 13.684 &   0.755 &   \dots &   0.472 &   0.969 &   1.658 &   2.217 &  0.01 \\
CS~29493-090 & 14.039 &   0.835 &   \dots &   0.516 &   1.028 &   1.740 &   2.366 &  0.03 \\
CS~29495-042 & 13.607 &   0.626 &   \dots &   0.415 &   0.825 &   1.261 &   1.737 &  0.03 \\
CS~29497-030 & 12.656 &   0.299 &   \dots &   0.215 &   0.440 &   0.694 &   0.911 &  0.02 \\
CS~29527-048 & 14.833 &   0.455 &   \dots &   0.314 &   0.632 &   1.046 &   1.356 &  0.02 \\
CS~29512-073 & 14.137 &   0.568 &   \dots &   0.377 &   0.753 &   1.235 &   1.624 &  0.05 \\
CS~30314-067 & 11.817 &   1.123 &   \dots &   0.642 &   1.234 &   2.058 &   2.789 &  0.07 \\
CS~31062-041 & 13.934 &   0.820 &   \dots &   0.458 &   0.950 &   1.545 &   2.113 &  0.03 \\
CS~31080-095 & 12.989 &   0.521 &$-$0.291 &   0.317 &   0.617 &   1.050 &   1.394 &  0.01
\enddata
\end{deluxetable*}

\subsubsection{\teff}

We used the color-{\teff} relations from \citet{alonso:96,alonso:99}.
Accurate photometry for our targets in Johnson $UBV$, and Kron-Cousins $RI$
was obtained as part of an ongoing program of observing metal-poor stars
\citep{beers:06a}. Near-IR $JHK$ magnitudes were retrieved from
the 2MASS catalog \citep{2MASS}. Table~\ref{Tab:Photometry} lists the
available photometry. We have adopted reddening values from
\citet{schlegel:98}. Almost all our targets have distances $> 1$\,kpc, much
larger than the 110\,pc scale height of the dust \citep{mendez:98}, hence the
assumption that all the reddening is between us and the star is appropriate.
For the more nearby stars, we calculated the correct amount of reddening to
use based on the model of \citet{mendez:98} (their Equation 4), iterating
until the following quantities converged: assumed reddening based on distance
from Sun, {\teff} from $V-K$, {\logg} and M$_V$ from isochrones, and finally
distance from M$_V$. This required between one and three cycles.

We adopt the $V-K$ temperature because it is independent of metallicity and
not strongly affected by carbon molecular features, contrary to the way that
bluer colors, such as $B-V$, have been shown to behave
\citep[e.g.][]{preston:01,cohen:06}.  A comparison of \teff $(V-K)$ with
temperatures derived from other colors shows an average offset of 
$\mteff (V-K) < \mteff (J-K)$ by
53\,K, $> \mteff (B-V)$ by 60\,K, $> \mteff (U-V)$ by 95\,K and $> \mteff
(V-I)$ by 74\,K. The higher temperatures for $V-K$ compared to $B-V$ and $U-V$
are expected if carbon absorption in the blue bands makes the stars appear
redder than the non C-rich calibration sample. A comparison with literature
values for our sample (Table~\ref{Tab:ModelAtmParLit}) shows that our {\teff}
estimates are hotter, on average, by 60\,K, with an rms scatter of 150\,K.  A
reasonable random error for our {\teff} determinations is therefore 150\,K.

\subsubsection{\logg}

We adopted the $Z=0.0001$ (${\rm [Fe/H]} = -2.31$, ${\rm [}\alpha{\rm /Fe]} =
+0.30$), 14\,Gyr isochrone from \citet{bergbusch:01}. Note that the age is
unimportant as long as it is greater than $\sim 10$\,Gyr. In choosing a
{\logg} for a given {\teff}, we assume that the star is a turnoff, subgiant,
or giant star, as appropriate for its location on the isochrone. In general,
this assumption is a good one. However, a few stars could be horizontal-branch
stars, and, indeed, a high-resolution analysis by \citet{mcwilliam:95b} showed
that CS~22947-187 has the lower gravity of a red horizontal-branch star,
$\log g = 1.30$, compared to our value of $\log g = 3.44$. We recalculated the
abundances using the {\logg} and $\xi$ value of \citet{mcwilliam:95b}. The
derived [Ca/H] did not change, while the $\log\epsilon ({\rm N})$ and the
$\log\epsilon ({\rm C})$ increased by 0.9\,dex and 0.80\,dex, respectively.
The [C/N] ratio, however, changed by a mere 0.10\,dex. Thus, for the small
subset of stars which we may mistakenly classify as subgiants rather than
horizontal-branch stars, the change in [C/N], the crucial aspect of this
paper, is smaller than that which could arise from other sources of error.

The major uncertainty in estimation of {\logg} for almost all stars is the uncertainty in
{\teff}, because a change of $\pm 150$\,K leads to a change of $\pm 0.14$\,dex
in {\logg} for a turnoff star, $0.19$\,dex for a subgiant and $0.35$\,dex for
a giant. A comparison with literature values (Table~\ref{Tab:ModelAtmParLit})
yields an average difference of $0.20$\,dex with an rms scatter of
$0.23$\,dex. Therefore, we adopt a {\logg} error of $0.3$\,dex for turnoff and
subgiant stars and $0.4$\,dex for giants, respectively.

\subsubsection{Microturbulent velocity}

Our abundances derived from lower dispersion data are not very sensitive to
$\xi$, the microturbulent velocity, but it still needs to be known to within
$1\,{\rm km\,s}^{-1}$. Fortunately, there exists a well-known correlation
between {\logg} and $\xi$. The HERES collaboration \citep{barklem:05} measured
$\xi$ for 254 metal-poor stars from high-resolution spectra. We fit their
{\logg} and $\xi$ values with a second-order polynomial and use the relation
for our sample:

\begin{equation}
\xi = 2.822 - 0.669\,\log g + 0.080\,(\log g)^2\;{\rm km}\,{\rm s}^{-1}
\end{equation}

\noindent This equation is valid for the range of surface gravities
exhibited by stars in the HERES sample $(1.0 < \log g < 4.2)$, which
encompasses the expected range of surface gravity for our program stars.  The
HERES data exhibit an rms scatter of 0.17\,km\,s$^{-1}$ around this relation.
We tested changes of $\pm 0.3$\,km\,s$^{-1}$ and found no change in our
synthesis; therefore uncertainties in $\xi$ do not contribute significantly to
our final abundance errors.

\subsubsection{[Fe/H]}

We require an estimate of [Fe/H] for two reasons, most importantly to have the
correct metallicity atmospheric model, and secondly in order to measure the
[C/Fe] ratio associated with the overall enhancement from AGB-star pollution.
For these purposes, an accuracy of $0.3$\,dex is sufficient. Test syntheses of
the NH and CH features and the \ion{Ca}{2}~K lines revealed only a small ($<
0.05$\,dex) dependence of these quantities on the metallicity of the model
atmosphere.  Furthermore, the carbon-enhanced nature of these stars has
already been established by larger-scale surveys with many stars to provide a
control sample. The lack of dependence on the overall metallicity of the model
is fortunate, as we did not use carbon-enhanced models, in keeping with most
of the previous work in the field. Masseron (2005, priv. comm.) has shown that
errors on the abundances from neglecting C-enhancement is mitigated in the
hotter, less C-rich stars that comprise our present sample.

We first measured [Ca/H] using the CaII~K line that is prominent in all of our
medium-resolution spectra, regardless of temperature and metallicity. The
linelist for this exercise was taken from \citet{castelli:03}. We then
converted to [Fe/H] using an assumed [Ca/Fe] ratio of $+0.30$\,dex, the
average [Ca/Fe] ratio determined by \citet{cayrel:04} in their sample of non
carbon-enhanced metal-poor stars.

\begin{deluxetable}{lccccc}
\tabletypesize{\footnotesize}
\tablenum{3}\label{Tab:ModelAtmPar}
\tablewidth{0pt}
\tablecaption{Model Atmosphere Parameters}
\tablehead{
\colhead{Star} &\colhead{E$_{B-V}$ (mag)} &\colhead {Teff (K)} & \colhead{logg} & 
\colhead{[m/H]} & \colhead{$\xi$ (km/s)} 
}
\startdata
BD$-18^{\circ}\,5550$     &  0.17  &   4806   &  1.72   & $-$2.89 & 1.91 \\
HD~122653      &  0.02  &   4615   &  1.27   & $-$2.47 & 2.10 \\
HD~160617      &  0.01  &   5882   &  3.69   & $-$1.69 & 1.45 \\
HD~186478      &  0.10  &   4831   &  1.78   & $-$2.63 & 1.89 \\
CS~22174-007   &  0.03  &   5059   &  2.37   & $-$2.00 & 1.69 \\
CS~22183-031   &  0.04  &   5196   &  2.76   & $-$2.79 & 1.59 \\
CS~22879-029   &  0.04  &   6300   &  3.90   & $-$1.93 & 1.43 \\
CS~22884-097   &  0.23  &   6460   &  4.00   & $-$1.94 & 1.43 \\
CS~22887-048   &  0.05  &   6455   &  3.99   & $-$2.22 & 1.43 \\
CS~22891-171   &  0.07  &   5297   &  3.07   & $-$2.45 & 1.52 \\
CS~22892-052   &  0.03  &   4861   &  1.86   & $-$2.76 & 1.86 \\
CS~22898-062   &  0.05  &   5309   &  3.10   & $-$1.74 & 1.52 \\
CS~22945-024   &  0.03  &   5289   &  3.04   & $-$2.26 & 1.53 \\
CS~22947-187   &  0.09  &   5489   &  3.44   & $-$2.25 & 1.47 \\
CS~22948-104   &  0.02  &   5270   &  2.99   & $-$2.39 & 1.54 \\
CS~22949-008   &  0.04  &   6144   &  3.82   & $-$1.92 & 1.44 \\
CS~22950-046   &  0.06  &   4604   &  1.25   & $-$3.29 & 2.11 \\
CS~22958-042   &  0.02  &   6224   &  3.86   & $-$2.65 & 1.44 \\
CS~22958-083   &  0.04  &   5189   &  2.74   & $-$2.50 & 1.59 \\
CS~22960-053   &  0.01  &   5061   &  2.38   & $-$3.08 & 1.68 \\
CS~22968-014   &  0.01  &   4892   &  1.93   & $-$3.30 & 1.83 \\
CS~29493-090   &  0.03  &   4739   &  1.56   & $-$2.82 & 1.97 \\
CS~29495-042   &  0.03  &   5400   &  3.32   & $-$2.30 & 1.49 \\
CS~29497-030   &  0.02  &   7163   &  4.20   & $-$2.20 & 1.43 \\
CS~29512-073   &  0.05  &   5751   &  3.62   & $-$2.10 & 1.45 \\
CS~30314-067   &  0.07  &   4476   &  0.96   & $-$2.67 & 2.25 \\
CS~31062-041   &  0.03  &   5042   &  2.32   & $-$2.30 & 1.70 \\
CS~31080-095   &  0.01  &   5972   &  3.73   & $-$2.75 & 1.44
\enddata
\end{deluxetable}

Several of our targets have been observed at high spectral resolution by
previous authors. These include some of the most C-rich stars in our sample,
as well as three bright metal-poor stars (HD~122563, HD~186478 and
BD$-18^{\circ}\,5550$) that we observed in order to compare with previous
results. In Table~\ref{Tab:ModelAtmParLit}, we compare our model atmosphere
parameters with values taken from the literature, restricting ourselves to
recent papers in the case of the HD and BD stars. In general, our model
atmospheres have somewhat higher {\teff} and higher {\logg} values. Our [Fe/H]
values are also higher, which is expected based on the differences in model
atmospheres.

In order to test this, and to show that our [Fe/H] values derived from
\ion{Ca}{2}~K are good to within about 0.15\,dex, we studied a subsample of
the stars that had equivalent widths (EWs) based on high-resolution spectra
available in the literature.  These EWs were run through our
analysis, adopting our model atmosphere parameters; the results are shown in
Table~\ref{Tab:MetallicityComparison}. The [Fe/H] derived from the CaII~K
line, with an assumed [Ca/Fe] of $+0.3$, is in good agreement with the [Fe/H]
derived from the EWs of \ion{Fe}{1} lines, except for CS~22183-031, whose
[Fe/H] in our analysis is 0.4\,dex higher than in the EW analysis. The
abundance ratios of [Ca/Fe], [Mg/Fe], [Ti/Fe] and [\ion{Fe}{2}/\ion{Fe}{1}]
are also quite reasonable, although our higher adopted gravity for
CS~22947-187 means that we derive much higher abundances for the ionized
species than for the neutral species. From
Table~\ref{Tab:MetallicityComparison}, we conclude that basing our model
atmosphere metallicities on the \ion{Ca}{2}~K line results in an offset of
$0.12$\,dex with a rms scatter of $\sim 0.16$\,dex, which is in line with the
expected uncertainty due to our fits of the \ion{Ca}{2}~K region. Our final
model atmosphere parameters are summarized in Table~\ref{Tab:ModelAtmPar}.

\begin{deluxetable*}{lrrrrl}
\tablenum{4}\label{Tab:ModelAtmParLit}
\tablewidth{0pt}
\tablecaption{Model Atmosphere Parameter Comparison}
\tablehead{\colhead{Star} & \colhead{\teff} & \colhead{log g} & \colhead{[m/H]}
& \colhead{$\xi$} & \colhead{Source} 
}
\startdata
BD$-18^{\circ}\,5550$   & 4750 & 1.4  &$-$3.06 & 1.80 & \citet{cayrel:04} \\
             & 4806 & 1.72 &$-$2.89 & 1.91 & This study \\
HD~122563    & 4600 & 1.10 &$-$2.82 & 2.00 &\citet{cayrel:04} \\
             & 4500 & 1.30 &$-$2.74 & 2.5  & \citet{westin:00} \\
             & 4615 & 1.27 &$-$2.47 & 2.10 & This study \\ 
HD~160617    & 5967 & 3.79 &$-$1.77 & 1.50 & \citet{jonsell:05b} \\ 
             & 5931 & 3.77 &$-$1.79 & 1.50 & \citet{akerman:04} \\
             & 5999 & 3.74 &$-$1.36 & 1.31 & \citet{gratton:00} \\
             & 5882 & 3.69 &$-$1.69 & 1.45 & This study \\
HD~186478    & 4700 & 1.30 &$-$2.59 & 2.00 &\citet{cayrel:04} \\
             & 4831 & 1.78 &$-$2.63 & 1.89 & This study \\
CS~22183-031 & 5270 & 2.8  &$-$2.93 & 1.20 & \citet{honda:04a} \\
             & 5196 & 2.76 &$-$2.79 & 1.59 & This study \\
CS~22892-052 & 4850 & 1.50 &$-$2.97 & 2.50 & \citet{norris:97b} \\
             & 4790 & 1.60 &$-$2.92 & 1.80 & \citet{honda:04a} \\
             & 4850 & 1.60 &$-$3.03 & 1.90 & \citet{cayrel:04} \\
             & 4760 & 1.30 &$-$3.10 & 2.29 & \citet{mcwilliam:95b} \\
             & 4861 & 1.86 &$-$2.76 & 1.86 & This study \\
CS~22947-187 & 5160 & 1.30 &$-$2.6  & 2.26 & \citet{mcwilliam:95b} \\
             & 5489 & 3.44 &$-$2.25 & 1.47 & This study \\
CS~22950-046 & 4640 & 0.85 &$-$3.5  & 2.68 & \citet{mcwilliam:95b} \\
             & 4730 & 1.30 &$-$3.30 & 2.02 & \citet{carretta:02} \\
             & 4604 & 1.25 &$-$3.29 & 2.11 & This study \\
CS~22958-042 & 6250 & 3.50 &$-$2.85 & 1.50 & \citet{sivarani:06} \\
             & 6224 & 3.86 &$-$2.65 & 1.44 & This study \\
CS~22968-014 & 4840 & 1.80 &$-$3.5  & 1.90 & \citet{mcwilliam:95b} \\
             & 4850 & 1.70 &$-$3.56 & 1.90 & \citet{cayrel:04} \\
             & 4892 & 1.93 &$-$3.30 & 1.83 & This study \\
CS~29497-030 & 6650 & 3.50 &$-$2.80 & 2.00 & \citet{sivarani:04a} \\
             & 7000 & 4.10 &$-$2.57 & 1.90 & \citet{ivans:05} \\
             & 7163 & 4.20 &$-$2.20 & 1.43 & This study \\
CS~30314-067 & 4400 & 0.70 &$-$2.85 & 2.50 & \citet{aoki:02b} \\
             & 4476 & 0.96 &$-$2.67 & 2.25 & This Study \\
CS~31085-090 & 6050 & 4.50 &$-$2.85 & 1.00 & \citet{sivarani:06} \\
             & 5972 & 3.73 &$-$2.75 & 1.44 & This study
\enddata
\end{deluxetable*}

\thispagestyle{empty}
\begin{deluxetable*}{lcccccccccc}
\tabletypesize{\scriptsize}
\tablenum{5}\label{Tab:MetallicityComparison}
\tablewidth{0pt}
\tablecaption{Metallicity Comparison}
\tablehead{\colhead{} & \multicolumn{2}{c}{Our Synthesis} & 
\multicolumn{7}{c}{Abundances Derived with Literature EWs and Our
Model Atmospheres} & \colhead{} \\
\colhead{Star} & \colhead{[\ion{Ca}{2}/H]} & 
\colhead{[Fe/H]} & \colhead{[\ion{Ca}{1}/H]} & \colhead{[\ion{Fe}{1}/H]}
& \colhead{[\ion{Ca}{1}/\ion{Fe}{1}]} & \colhead{[\ion{Mg}{1}/\ion{Fe}{1}]} &
\colhead{[\ion{Ti}{1}/\ion{Fe}{1}]} & \colhead{[\ion{Ti}{2}/\ion{Fe}{1}]} & 
\colhead{[\ion{Fe}{2}/\ion{Fe}{1}]} & \colhead{EW Source} 
}
\startdata
CS~22183-031 & $-$2.49 &$-$2.79 &$-$2.76 &$-$3.19 & 0.43 & 0.72 & 0.68 & 0.58 & $+$0.13  & 1\\
CS~22947-187 & $-$1.95 &$-$2.25 &$-$1.77 &$-$2.11 & 0.34 & 0.44 & 0.30 & 0.96 & $+$0.59 & 2\\
CS~22892-052 & $-$2.46 &$-$2.76 &$-$2.65 &$-$2.94 & 0.30 & 0.46 & 0.20 & 0.16 & $+$0.00 & 3\\
CS~22950-046 & $-$2.99 &$-$3.29 &$-$3.23 &$-$3.46 & 0.23 & 0.63 & 0.19 & 0.22 & $+$0.00 & 4\\
CS~22968-014 & $-$3.00 &$-$3.30 &$-$3.35 &$-$3.38 & 0.03 & 0.47 & 0.10 & 0.06 &$-$0.05 & 5\\
CS~29497-030 & $-$1.90 &$-$2.20 &$-$1.85 &$-$2.30 & 0.45 & 0.66 & 0.65 & 0.47 &$-$0.15 & 6\\
CS~30314-067 & $-$2.37 &$-$2.67 &$-$2.61 &$-$2.71 & 0.10 & 0.69 & 0.22 & 0.42 &$-$0.06 & 7
\enddata
\tablerefs{(1)\citet{honda:04a}, (2) \citet{mcwilliam:95a}, (3) \citet{sneden:03}, (4)\cite{carretta:02},
(5) \citet{cayrel:04}, (6) \citet{ivans:05}, (7)\citet{aoki:02a}}
\end{deluxetable*}

\subsection{C and N Measurements and Upper Limits}

\begin{figure}
\includegraphics[angle=270,width=8cm]{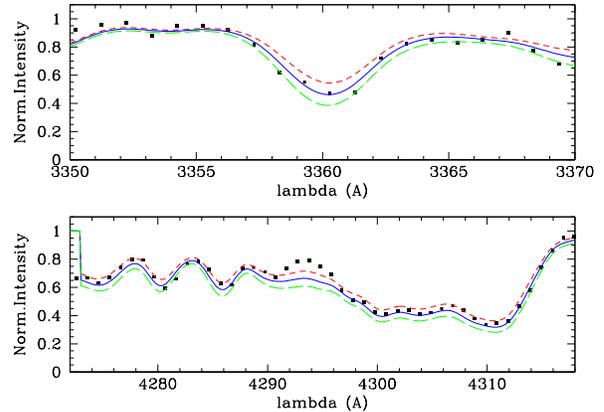} 
\caption{\label{Fig:CS22891}Example of syntheses for CS~22891-171
for (top) NH and (bottom) CH. In each case, the solid line represents
the adopted abundance, the dashed lines $\pm0.2$ dex of the adopted
abundance. The CS~22891-171 spectrum has a S/N per resolution element of 58,
about average for our spectra. }
\end{figure}

We measured C and N abundances by creating a synthetic spectrum for each star
and comparing it with the observed spectra (Figure~\ref{Fig:CS22891}). When the NH and CH regions were
synthesized, the abundance ratios of [Mg/Fe], [Ca/Fe] and [Ti/Fe] were set to
be $+0.3$, $+0.3$, and $+0.2$, based on the \citet{cayrel:04} results for the
enhancements in the $\alpha$-elements for metal-poor halo stars. Other
elemental abundance ratios may also differ from solar in these stars, but
their lines are either not strong or sufficiently plentiful to noticeably
affect the synthesis.  We calculated the molecular equilibrium among the
species H$_2$, CH, NH, OH, C$_2$, CN, CO, N$_2$, NO, O$_2$, H$_2$O, and
CO$_2$. Therefore, our derived C and N abundances could depend on the
adopted O abundance. We set the [O/Fe] at $+0.4$ \citep{fulbright:99} in our
syntheses, however the C and N abundances are not sensitive to the adopted
[O/Fe] unless it exceeds $+1.5$, and then only at the $0.01$--$0.02$\,dex level.
The synthesis of the NH band does not depend on the C abundance, which was one
of the advantages of this program. The random uncertainty from continuum
placement, $S/N$, and imperfect line lists is estimated to be $0.2$\,dex for
the C, N and Ca syntheses based on by-eye fits to the data with different
abundances and continuum placements. 

Our resolution is sufficiently low that a determination of a lower limit on N
for some of our program stars relies on the lowest point in the spectra in the
region of the NH feature. We found a lower limit by selecting the lowest pixel
value within a window $0.8$\,{\AA} wide, centered on the expected position of the
maximum N absorption. The rms scatter was determined from the $S/N$, and a
4-$\sigma$ offset downward was added to that data point. Then, the lowest points
in many NH syntheses were found, and interpolated in order to yield a lower
limit on the N abundance.

\subsection{Three-Dimensional and Non-LTE Effects}

Three-dimensonial effects are likely to be very important in deriving correct
C and N abundances. Preliminary calculations in red giant atmospheres for the
NH lines at 3360\,{\AA} and the CH lines at 4305\,{\AA} indicate that the
corrections can be as large as $0.4$ and $0.5$\,dex, respectively
(Garcia-Perez, priv. comm.) and 0.8 dex and larger for the
most iron-poor stars \citep[e.g.][]{collet:06}. Because we are
primarily concerned with [C/N] ratios in this paper, our results are not as
sensitive to these corrections, and we ignore them for now. If 3-D corrections
of this magnitude are applied to our [C/Fe] values, in many cases the values
will drop below ${\rm [C/Fe]} = 0$, and one might wonder if we still have a
useful sample. However, in that case, [C/Fe] for metal-poor field stars in
general will also drop by the same amount, and our stars will still be
enriched {\it relative} to normal stars. Hence the assumption that
C-enrichment is due to AGB stars still would stand.

\subsection{Error Analysis}

There are two major sources of error that we need to quantify. The first is
the choice of molecular parameters, including the absolute oscillator
strengths and dissociation potentials, which were discussed in
\S\,\ref{Sect:AtomicData}. The second is the choice of model atmosphere
parameters, in particular {\teff}. Our samples includes stars with a wide
range of effective temperature and gravity, therefore we selected three stars
covering these ranges, CS~22887-042, CS~22947-187, and CS~29493-090, for the
error analysis. The changes in $\log\epsilon$ for C, N and Ca are listed in
Table~\ref{Tab:DeltaLogEps}. The dependence on $\xi$ and metallicity are much
smaller than those for {\teff} and {\logg}, and hence will be ignored. The
increase in $\log\epsilon$ with increasing temperature is offset by a decrease
with increasing gravity when the actual model atmosphere parameters are used.
To account for the expected correlations between error sources, we adopt the
following equation:

\begin{deluxetable*}{lrrrrrr}
\tablenum{6}\label{Tab:DeltaLogEps}
\tablewidth{0pt}
\tablecaption{$\Delta$ Abundance for C, N, and Ca with Model Atmosphere Changes}
\tablehead{\colhead{Star} & \multicolumn{2}{c}{$\Delta$log$\epsilon$(C)} & 
\multicolumn{2}{c}{$\Delta$log$\epsilon$(N)}  & 
\multicolumn{2}{c}{$\Delta$log$\epsilon$(Ca)} \\
\colhead{} &
\colhead{$\Delta$ \teff} & \colhead{$\Delta$\logg} &
\colhead{$\Delta$ \teff} & \colhead{$\Delta$\logg} &
\colhead{$\Delta$ \teff} & \colhead{$\Delta$\logg} \\
\colhead{} &
\colhead{$+150$\,K } & \colhead{$+0.3$\,dex} &
\colhead{$+150$\,K } & \colhead{$+0.3$\,dex} &
\colhead{$+150$\,K } & \colhead{$+0.3$\,dex} 
}
\startdata
CS~22887-048 & 0.250 & $-$0.125 & 0.325 & $-$0.125 & 0.125 & 0.000 \\ 
CS~22947-187 & 0.275 & $-$0.100 & 0.275 & $-$0.125 & 0.015 & $-$0.050 \\
CS~29493-090 & 0.310 & $-$0.125 & 0.300 & $-$0.225 & 0.100 & $-$0.025
\enddata
\end{deluxetable*}

{\setlength
\arraycolsep{1pt}
\begin{eqnarray}
\sigma^2_{log\epsilon}= \sigma^2_{syn} +
\left({\partial log \epsilon\over\partial T}\right)^2 
\sigma^2_T + 
\left({\partial log\epsilon\over\partial \mlogg}\right)^2 
\sigma^2_{\mlogg} 
{}
\nonumber\\
{}
+  2\left({\partial log\epsilon\over\partial T}\right)
\left({\partial log\epsilon
\over\partial \mlogg}\right)\sigma_{T\mlogg} 
\end{eqnarray}}
\noindent where $\sigma_{syn}$ is the abundance error associated with
the synthesis and is 0.2 dex. In this case, where \logg{} is
derived from \teff, we write $\sigma_{T \mlogg}$ as

\begin{equation}
\sigma_{T \mlogg}=\left(\partial \mlogg \over \partial T \right)\sigma^2_T
\end{equation}

The slope of the {\logg}-{\teff} relation depends on the evolutionary
state of the star. We found slopes of $2.33\times 10^{-3}$ for the giant
CS~29493-090, $1.27\times 10^{-3}$ for the subgiant CS~22947-187,
and $9.33\times 10^{-4}$ for the turnoff star CS~29493-090.
The errors in [X/H] were calculated using equation 2, while the
the abundance ratio errors for [C/N], [C/Fe] and [N/Fe] were calculated
using a modification of Equations A19 and A20 from \citet{mcwilliam:95b}. Our
uncertainties for turnoff, subgiant, and giant stars are summarized
in Table~\ref{Tab:ErrorSummary}.


\begin{deluxetable*}{lccccccc}
\tabletypesize{\footnotesize}
\tablenum{7}\label{Tab:ErrorSummary}
\tablewidth{0pt}
\tablecaption{Summary of Errors}
\tablehead{\colhead{Star} & \colhead{{\teff} range} &
\colhead{$\sigma$(log$\epsilon$(C))} & \colhead{$\sigma$(log$\epsilon$(N))} &
\colhead{$\sigma$(log$\epsilon$(Fe))} &  
\colhead{$\sigma$([C/Fe])} & \colhead{$\sigma$([N/Fe])} &
\colhead{$\sigma$([C/N])} 
}
\startdata
turnoff & \teff $> 6400$K & 0.30 & 0.36 & 0.24 & 0.31 & 0.35 & 0.31 \\
subgiant & $5400$K$ <$\teff$<6400$K & 0.30 & 0.31 & 0.24 & 0.30 & 0.32 & 0.30 \\
giant & \teff $<5400$K & 0.25 & 0.25 & 0.21 & 0.29 & 0.25 & 0.41
\enddata
\end{deluxetable*}

\section{Results}\label{Sect:Results}

\begin{figure}
\includegraphics[width=8cm]{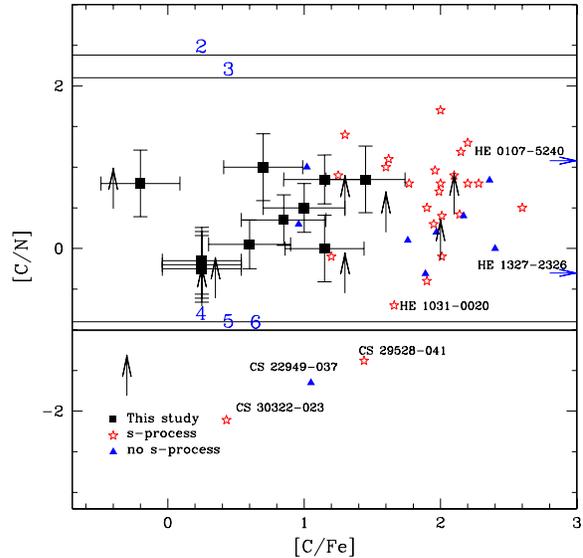}
\caption{\label{Fig:CFeCNnew} [C/N] vs. [C/Fe] for our sample of stars (filled 
  squares) as well as literature stars with (stars) and without (triangles)
  s-process enhancements. The lines are the same as in Figure~\ref{Fig:CFeCNold}.
  The [C/N] ratios for the most iron-poor stars are indicated by rightward
  pointing arrows because their [C/Fe] ratios (HE~0107$-$5240, ${\rm [C/Fe]} =
  +3.71$, and HE~1327$-$2326, ${\rm [C/Fe]} = +4.26$) are off the scale of the
  plot.}
\end{figure}

The abundances of N and C, derived from the NH, CN, and CH lines, are listed
in Table~\ref{Tab:LogEpsNC} and plotted in Figure~\ref{Fig:CFeCNnew}. Twenty-one
of the stars in our sample are C-mild to C-rich and are pertinent to the
problem of the ``missing'' NEMP stars. The seven often-observed calibration
stars (CS~22892-052, CS~22968-014, CS~22950-046 and the HD/BD stars) are not
shown in Figure~\ref{Fig:CFeCNnew}. It is clear from the low {\teff} of
CS~22950-046 that its low [C/N] ratio is due to internal mixing
\citep{gratton:00,spite:05}, and it is similar to the other low {\teff}
calibration giants, HD~122563 and HD~186478.

Two facts are immediately apparent from inspection of
Figure~\ref{Fig:CFeCNnew}. (1) We have mild C-enhancements for many stars in
our sample, and have managed to fill in the $+0.5 < {\rm [C/Fe]} < +1.0$
region of the diagram and (2) our [C/N] values are not consistent with those
expected to be produced by HBB in AGB stars, and indeed are rather similar to
previous results. Therefore, it appears that an observational bias is not the
solution to the scarcity of NEMP stars.

Table~\ref{Tab:LogEpsNCcomp} lists our derived [C/Fe] and [C/N] ratios
compared with available literature values for stars in common, while
Table~\ref{Tab:LogEpsNCLit} lists the additional literature values used in
Figure~\ref{Fig:CFeCNnew}. There are three stars in the non-calibration sample
with ${\rm [C/Fe]} < 0$. However, examination of Table~\ref{Tab:LogEpsNCcomp}
shows that our [C/Fe] ratios are in general lower than the literature values.
Because these stars were chosen out of a large sample with consistent
C-abundance determinations \citep{rossi:05}, we believe they are still
C-enhanced relative to the field population.

There are four stars with literature [C/N] values in the interesting low range
of [C/N] in Figure~\ref{Fig:CFeCNnew}; these deserve some special discussion.
Based on the observations in Figure~\ref{Fig:CFeCNnew}, we define the
class of NEMP stars as having [N/Fe]$>$0.5 and [C/N]$<-$0.5. This marks stars
that have distinctly lower [C/N] ratios than the rest of the CEMP stars, 
although we note that to agree with the predictions of HBB, these stars
would need to have [C/N] $\sim -1$.
Two of these, CS~29528-041 and HE~1031$-$0020 have properties similar to the
NEMP stars we were expecting to find in much greater numbers. Of the other
two, CS~22949-037 is not s-process rich (but it is oxygen rich), and
CS~30322-023 is probably an intrinsic thermally pulsing AGB star
\citep{masseron:06}. The statistics for the presence of NEMP stars in
the literature confirm the rarity that led to the lack of them in our
sample.

\begin{itemize}
  
\item {\bf CS~29528-041:} \citet{sivarani:06} report on the elemental
  abundances of this C-rich, but even more N-rich star. It has the expected
  abundance ratios for NEMP stars: ${\rm [Fe/H]} = -3.3$, ${\rm [C/Fe]} =
  +1.6$, ${\rm [N/Fe]} = +3.0$, hence ${\rm [C/N]} = -1.4$.  Interestingly,
  this star also has detectable lithium, albeit at a value well below the
  Spite Plateau, $A({\rm Li}) = 1.7$. It is s-process-element rich as well,
  with ${\rm [Ba/Fe]} = +1.0$, which indicates that more massive AGB stars can
  produce the s-process.
\item {\bf HE~1031$-$0020:} \citet{cohen:06} found a [C/N] of $-0.85$ for this
  s-process-element rich star, which is one of the less C-enhanced stars in
  their sample, with ${\rm [C/Fe]} = +1.6$.
\item {\bf CS~22949-037:} \citet{norris:02} showed that this star has a [C/N]
  ratio of $-1.25$, while \citet{depagne:02} found an extremely high [O/Fe]
  ratio of $+1.97$. Models of EMP or $Z=0$ low-mass AGB stars based on
  standard mixing assumptions (no convective overshooting at bottom of
  He-shell flash convection zone) do predict a range of O overabundances
  depending on mass \citep{siess:02,herwig:04b}. For $Z=0.0001$, ${\rm [O/Fe]}
  $ can be up to 1.7 dex for low-mass models. Like C and N, the O production in AGB stars is
  primary, and therefore larger overabundances are expected with low metal
  content. However, the large N overabundance observed in this star is only
  predicted for more massive AGB stars for which the O overabundance is
  probably smaller. This, and the lack of enrichment in the s-process
  elements, supports the hypernova hypothesis for this star.
\item {\bf CS~30322-023:} \citet{masseron:06} have argued that this extremely
  low metallicity star, with ${\rm [Fe/H]} = -3.5$, and with a clear
  s-process-element signature, is likely to be an example of an {\it
    intrinsic} AGB star, caught during the brief stage of evolution when
  thermal pulses are occurring.  This star is mildly carbon-rich, with ${\rm
    [C/Fe]} = +0.6$, while the nitrogen abundance ratio is quite high, ${\rm
    [N/Fe]} = +2.8$. We would not classify this an NEMP star, however, if it
  is an intrinsic AGB star, because there is no binary mass transfer, in
  analogy to the CEMP-s stars. Note, however, that if the identification of
  this star as a TP-AGB is correct, its mass cannot be high, but rather, it
  must be on the order of $0.8\,{\rm M}_{\odot}$. Since at masses this
low, HBB is not expected to occur, \citet{masseron:06} suggest an unknown
mixing process that happens in VMP stars is responsible.
\end{itemize}

\begin{deluxetable*}{lrrrrrrr}
\tabletypesize{\footnotesize}
\tablenum{8}\label{Tab:LogEpsNC}
\tablewidth{0pt}
\tablecaption{C, N, and [Fe/H] for Our Sample}
\tablehead{\colhead{Star} & \colhead{log$\epsilon$(C)} 
& \colhead{[C/Fe]} & \colhead{log$\epsilon$(N)} & \colhead{[N/Fe]}
& \colhead{[Fe/H]}
 & \colhead{[C/N]}
}
\startdata
BD$-18^{\circ}\,5550$   & 5.47&    $-$0.20 & \nodata  &   \nodata  & $-$2.89 & \nodata\\
HD~122563    & 5.14& $-$0.95    & 6.58     &   $+$1.00   &  $-$2.47 &$-$1.95 \\
HD~160617    & 6.27 &  $-$0.60    & 6.76     &  $+$0.40    & $-$1.69 & $-$1.00\\
HD~186478    & 5.73&   $-$0.20  & 6.12     &  $+$0.70    &  $-$2.63 & $-$0.90\\
CS~22174-007 & 6.16 &  $-$0.40 & $<$5.16  &  $<-$0.89& $-$2.00 &$>+$0.49   \\
CS~22183-031 & 5.47 &  $-$0.30 & $<$6.77  &  $<+$1.51   & $-$2.79 &$>-$1.81  \\
CS~22879-029 & 7.93 &   $+$1.30 &$<$7.02  &  $<+$0.90 & $-$1.93 &$>+$0.40   \\
CS~22884-097 & 7.92 &    $+$1.30 & $<$7.96  & $<+$1.85    & $-$1.94 &$>-$0.55   \\
CS~22887-048 & 7.84 &    $+$0.85 &  6.98  &    $+$0.50  & $-$2.79& $+$0.35 \\
CS~22891-171 & 7.56 &     $+$1.45 &   6.20  &    $+$0.60  & $-$2.45 &$+$0.85  \\
CS~22892-052 & 6.40&     $+$0.60 & 5.14  & $-$0.15  & $-$2.76 &$+$0.75  \\
CS~22898-062 & 6.62 & $-$0.20  &  5.31  & $-$1.00  & $-$1.74 &$+$0.80  \\
CS~22945-024 & 7.00 &    $+$0.70  &  5.49 & $-$0.30  & $-$2.26 &$+$1.00  \\
CS~22947-187 & 6.91 &    $+$0.60  &  6.35  &    $+$0.55  & $-$2.25 &$+$0.05  \\ 
CS~22948-104 & 6.42 &     $+$0.25 &$<$6.62  &  $<+$0.96      & $-$2.39 &$>-$0.71 \\
CS~22949-008 & 7.79 &    $+$1.15  &  6.43  &    $+$0.30  & $-$1.92 &$+$0.85  \\
CS~22950-046 & 4.77 &   $-$0.50 &   6.16  &    $+$0.30  & $-$3.29 &$+$0.80 \\
CS~22958-042 & 8.31 &     $+$2.40 & $<$7.29  & $<+$1.89  & $-$2.65 &$>+$0.51   \\
CS~22958-083 & 6.41   &     $+$0.35 &$<$6.51  &  $<+$0.96  & $-$2.50 &$>-$0.61   \\
CS~22960-053 & 6.63 &    $+$1.15  &  6.12  &    $+$1.15  & $-$3.08 &$+$0.00  \\
CS~22968-014 & 5.36   &   $+$0.10 &$<$4.88  & $<+$0.13  & $-$3.30 &$>-$0.03   \\
CS~29493-090 & 5.99 &    $+$0.25  &  5.68  &    $+$0.45  & $-$2.82 &$-$0.20  \\
CS~29495-042 & 7.26 &    $+$1.00  &  6.25  &    $+$0.50  & $-$2.30 &$+$0.50  \\
CS~29497-030 & 8.36 &   $+$2.00   &$<$7.99  & $<+$2.14  & $-$2.20 &$>-$0.14  \\
CS~29512-073 & 7.51 &     $+$1.05 &5.95  &    $+$0.00  & $-$2.10 &$+$1.05  \\
CS~30314-067 & 6.14 &    $+$0.25  &  5.88  &    $+$0.50  & $-$2.67 &$-$0.25   \\
CS~31062-041 & 6.51 &    $+$0.25  & 6.15  &    $+$0.40  & $-$2.30 &$+$0.15   \\
CS~31080-095 & 7.91 &     $+$2.10 & $<$6.56  & $<$ 1.26 & $-$2.75 &$>+$0.84
\enddata
\end{deluxetable*}

\begin{deluxetable*}{lrrrrl}
\tablenum{9}\label{Tab:LogEpsNCcomp}
\tablecaption{Comparison with Literature Values}
\tablehead{
\colhead{Star} & \colhead{[C/H]} & \colhead{[N/H]} & \colhead{[C/N]} & \colhead{[C/Fe]} & \colhead{Source} 
}
\startdata
BD$-18^{\circ}\,5550$ & $-$3.08&  $-$3.02& \nodata& $-$0.02 & \citet{spite:05} \\
            & $-$3.09 & \nodata & \nodata  & $-$0.20    & This study \\
HD~122563 & $-$3.29 & $-$1.72 & $-$1.57 &$-$0.47  & \citet{spite:05} \\
          &$-$3.20 &  $-$1.65 &$-$1.55& $-$0.46 & \citet{westin:00} \\
          &  $-$3.42  & $-$1.47& $-$1.95 & $-$0.95 & This study \\
HD~160617 & $-$1.82 &\nodata& \nodata&   $+$0.03 & \citet{akerman:04} \\ 
          & $\leq -$1.69 & $-$0.34 & $\leq -1.35$& $\leq+ 0.30$ & \citet{laird:85}\\ 
          & $-$2.29 &  $-$1.29 & $-$1.00 &  $-$0.60  & This study \\
HD~186478 & $-$2.89 & $-$1.57 & $-$1.32&$-$0.30 &  \citet{spite:05} \\
          & $-$2.83  &  $-$1.93& $-$1.00 & $-$0.20 & This study \\
CS~22892-052 & $-$2.06 & \nodata & \nodata & $+$0.98 & \citet{mcwilliam:95b} \\
             & $-$2.26 & $-$2.22 & $-$0.04 & $+$0.88 & \citet{sneden:03} \\
             & $-$2.16 & $-$2.91 & $+$0.75 & $+$0.60 & This study \\
CS~22958-042 & $+$0.14 & $-$1.00 & $+$1.14 & $+$3.01 & \citet{sivarani:06} \\
             & $-$0.25 & $<-0.76$ & $>+$0.51 & $+$2.40 & This study \\
CS~30314-067 & $-$2.4 & $-$1.7 & $-$0.70 & $+$0.50 & \citet{aoki:02a} \\
             & $-$2.42 & $-$2.17 &  $-$0.25 & $+$0.25 & This study \\
CS~31085-095 & $-$0.31 & $-$2.40 & $+$2.09 & $+$2.56 & \citet{sivarani:06} \\
             & $-$0.65 &$-$1.49  & $>+$0.84 & $+$2.10 & This study
\enddata
\end{deluxetable*}

\begin{deluxetable*}{llrrrlc}
\tablenum{10}\label{Tab:LogEpsNCLit}
\tablecaption{Additional Literature Values}
\tablehead{\colhead{Star} & \colhead{log$\epsilon$(C)} & \colhead{log$\epsilon$(N)} & \colhead{[Fe/H]}
 & \colhead{[C/N]} & \colhead{Source}  & \colhead{s-process?}
}
\startdata
\multicolumn{7}{c}{Original Literature Sample}  \\
CS~22877-001 & 6.86 & 5.35 &$-$2.72 &$+$1.00 & \citet{aoki:02a}  & no \\
CS~22880-074 & 7.93 & 6.02 &$-$1.93 &$+$1.40 & \citet{aoki:02b} &  yes \\
CS~22881-036 & 8.46 &  6.99 &$-$2.06 &$+$0.96 & \citet{preston:01} & yes \\
CS~22898-027 & 8.51 & 6.70  &$-$2.25 &$+$1.30 & \citet{aoki:02b}& yes \\
CS~22942-019 & 7.92 & 6.01  &$-$2.64 &$+$1.70 & \citet{aoki:02b} &  yes\\
CS~22948-027 & 7.96 & 7.25 &$-$2.57 &$+$0.20 & \citet{aoki:02a} & no \\
CS~22957-027 & 7.80 & 6.45 &$-$3.38 &$+$0.84 & \citet{aoki:02l} & no\\
CS~29497-034 & 7.56 & 7.45 &$-$2.90 &$-$0.40 & \citet{hill:00} & yes \\
CS~29498-043 & 6.70 & 6.50 &$-$3.75 &$-$0.31 & \citet{aoki:02l} & no \\
CS~29502-092 & 6.76 & 5.95 &$-$2.76 &$+$0.30 & \citet{aoki:02a}  & no \\
CS~29526-110 & 8.38 & 7.07  &$-$2.38 &$+$0.80 & \citet{aoki:02b}&  yes\\
CS~30301-015 & 7.52 & 6.01  &$-$2.64 &$+$1.00 & \citet{aoki:02b} &  yes\\
CS~31062-012 & 8.11 & 6.70  &$-$2.55 &$+$0.90 & \citet{aoki:02b} &  yes\\
CS~31062-050 & 8.24 & 6.93  &$-$2.32 &$+$0.80 & \citet{aoki:02b} &  yes\\
HD~196944    & 7.51 & 7.10  &$-$2.25 &$-$0.10 & \citet{aoki:02b} &  yes\\ 
HE~0024-2523 & 8.44 & 7.43 & $-$2.72 &$+$0.50 & \citet{lucatello:02} & yes \\
LP~625-44 & 8.00 & 6.30   & $-$2.71 & $+$1.19& \citet{aoki:01} & yes \\
LP~706-7 & 7.96 & 7.03    & $-$2.74 & $+$0.42& \citet{aoki:01} & yes \\
\multicolumn{7}{c}{New and Updated Literature Sample} \\
HE~0012-1441 &7.66 & 6.05 &$-$2.52 &$+$1.10 & \citet{cohen:06} & yes\\ 
HE~0058-0244 &7.76 & 6.95 &$-$2.75 &$+$0.30 & \citet{cohen:06} &yes \\ 
HE~0107-5240 & 6.81 & 5.22 & $-$5.46 &$+$1.08 & \citet{christlieb:04} & no \\
HE~0143-0441 & 8.26 & 7.35 &$-$2.31 &$+$0.40 & \citet{cohen:06} & yes\\ 
HE~0212-0557 & 8.06 & 6.75 &$-$2.27 &$+$0.80 & \citet{cohen:06} & yes\\ 
HE~0336+0113 & 8.16 & 6.85 &$-$2.68 &$+$0.80 & \citet{cohen:06} & yes\\ 
HE~1150-0428 & 7.66 & 7.15 &$-$3.30 & $+$0.00& \citet{cohen:06} & no \\ 
HE~1410+0213 & 8.16 & 6.85 &$-$2.16 & $+$0.10& \citet{cohen:06} & no\\ 
HE~1434-1442 & 8.16 & 6.95 &$-$2.39 &$+$0.70& \citet{cohen:06} & yes\\ 
HE~1509-0806 & 7.66 & 7.25 &$-$2.91 & $-$0.10& \citet{cohen:06} & yes\\ 
HE~2158-0348 & 7.76 & 6.75 &$-$2.70 & $+$0.50& \citet{cohen:06} & yes\\ 
HE~2232-0603 & 7.96 & 6.55 &$-$1.85 & $+$0.90& \citet{cohen:06} & yes\\ 
HE~2356-0410 & 7.66 & 6.75 &$-$3.07 &$+$0.40 & \citet{cohen:06} & no\\ 
HE~1327-2326 & 6.99 & 6.83 & $-$5.73 &$-$0.37 &\citet{aoki:06} & no \\
\multicolumn{7}{c}{Nitrogen-Rich Stars} \\
CS~29529-041 & 6.70 & 7.57 & $-$3.32 & $-$1.38 & \citet{sivarani:06} & yes \\
CS~22949-037 & 5.82 & 6.96 & $-$3.79 & $-$1.65 & \citet{norris:02} & no \\
CS~30322-023 & 5.60 & 7.20 & $-$3.39 & $-$2.11 & \citet{masseron:06} & yes \\
HE~1031-0020 & 7.36 & 7.55 &$-$2.86 &$-$0.70 & \citet{cohen:06} & yes\\ 
\enddata
\end{deluxetable*}

\section{Discussion: Why the Lack of Observed NEMP Stars?}\label{Sect:Discussion}

We expected to find that between 12\,\% and 35\,\% of our sample would turn
out to be NEMP stars (see \S\,\ref{Sect:IMAGBpollution}), depending on our assumption concerning the mass ratios in
the binary progenitors. Instead, we found no candidates in our sample of 21
stars, and only two in the recent literature. The presence of these two ${\rm
  [C/N]}\sim -1.00$, s-process-rich stars suggest that NEMP stars do in fact
exist, and their abundance patterns are as expected from the models, just not
in the predicted numbers that should be found. However, there remain several
outstanding questions concerning the nature of our sample which we should
examine before asking if models of primary nitrogen production by
intermediate-mass AGB stars are incorrect.

\subsection{Are we observing stars that have been polluted by AGB 
 stars?}\label{Sect:AGBpollution}

AGB stars are by far the most likely culprits for the pollution of our sample.
The majority of CEMP stars are CEMP-s stars, and much observational evidence,
reviewed in the Introduction, points to their being the result of AGB mass
transfer. However, it is possible that stars with lower [C/Fe] values have a
smaller percentage that are s-process-rich than those with ${\rm [C/Fe]} >
1.0$. We have information on the s-process abundances for four stars in the
sample: CS~22947-187 and CS~31085-090 are s-process-rich, while CS~30314-067
and CS~22958-042 are not and there appears to be no correlation with
C-richness.  We are currently obtaining follow-up high-resolution spectra for
many of stars in our sample to check for radial-velocity variations and
s-process enhancements.

\subsection{Are we observing stars that have been polluted by
  intermediate-mass AGB stars?}\label{Sect:IMAGBpollution}

Although we are confident that most of our stars have been polluted by AGB
stars, we have considerably less information on the distribution of masses of
those AGB stars. Our initial estimate was based on the mass ratio ($q$)
distribution deduced by \citet{pinn:06} from a variety of studies,
particularly the data on massive binaries in the Small Magellanic Cloud from
\citet{harries:03} and \citet{hilditch:05}. Their most likely distribution of $q$
fractions is that 45\,\% of binary systems are ``twins'', with the primary and
secondary masses the same to within 5\,\%, while the $q$ of the other 55\,\%
of systems are matched by a flat mass distribution. We assumed that the
observed stars, being main-sequence turnoff, subgiant, giant or horizontal
branch stars, all have main-sequence masses close to 0.9\,{\solmass}.
Therefore, their ``twin'' AGB stars would be $< 1$\,{\solmass} and would have
created high [C/N] ratios. Therefore, the NEMP stars must come from the flat
mass distribution. We adopt 3.5\,{\solmass} as the dividing line between those
AGB stars that undergo HBB and those that do not. Finally, the maximum mass of
an AGB star at ${\rm [Fe/H]}=-2.3$ is 7.5\,{\solmass} (Poelarends et al., in
prep).
Combining all of this information, we find that there should be roughly 33\,\% NEMP stars
and 67\,\% CEMP-s stars contained in our sample.

There may be difficulties with our predictions.  For example, while \citet{pinn:06} consider the mass ratios using studies that
cover a wide range of masses, none of the studies is able to cover a wide
range of $q$. The SMC binary studies only reach a mass ratio of
$\sim 0.4$, which is larger than minimum $q$ we are interested in, 0.12, and
it is conceivable that the distribution will change as the mass difference
between the two components becomes more extreme. Such studies are difficult
because of the different luminosities and timescales of the two kinds of
stars, but we do have some information on the $q$ distribution of binaries
with solar-type primaries \citep{duquennoy:91}. For the longer period ($P >
3000$ days) binaries, the data suggested a secondary star drawn from an
IMF-weighted distribution. However, for the shorter period binaries,
\citet{mazeh:92} found a distribution that was much flatter, with perhaps a
rise toward $q=1$. For lack of other information, if we make the assumption
that the mass of the companion of the observed stars is drawn from a \citet{salpeter:55} mass function between 1\solmass{} and 7.5 \solmass, we find that there should be 1 NEMP for every 6.9 CEMP-s stars.
This is a much smaller ratio, but still not small enough to be supported by
the observational data. All of these numbers were obtained assuming that there
was not a bias towards finding NEMP stars in the C-mild sample; if such a bias
existed, the disagreement with theory would only be strengthened.

\subsection{Could binarity be affecting our results?}

The fact that these AGB stars were in binary systems may
have affected their nucleosythesis yields by increasing their mass-loss rates
and ending the production of N by HBB.
This could be due to tidal synchronization between rotational and orbital
motions, as observationally demonstrated for RGB stars by
\citet{demedeiros:02}. Even after HBB is terminated, dredge-up
can continue if the minimum envelope mass for dredge-up is less than that for
HBB, and the resulting [C/N] ratio may be altered. \citet{frost:98} argued
that the presence of high- luminosity C stars in the Large Magellanic Cloud
could be explained by continued dredge-up after HBB has ceased.  It is not
clear if this same reasoning would apply to our present sample, which has much
lower metallicity. However, if tidal synchronization leads to a spin-up
mixing, N production should be enhanced, possibly balancing the effect of
enhanced mass loss. Without detailed models we can only speculate, but it is
not immediately clear that binarity could inhibit N production at the bottom
of more massive AGB stars.  

\subsection{Could mass ratios affect mass transfer?}

\citet{jorissen:92b} summarize the situation regarding mass transfer in both
Roche-lobe overflow and wind models. For large mass ratios, only wind models
were stable, because the large mass-transfer rates in overflow 
scenarios puff up the secondary star
and lead to a contact binary situation \citep{tout:91}.  \citet{vanture:92b}
argued that the mass transfer from a wind would not be sufficient to create CH
stars, and therefore only in binary systems with Roche-lobe overflow would the
nucleosynthesis of the AGB star be preserved in the still-shining secondary.
Much work remains to be done in understanding mass-transfer scenarios, but if
NEMP stars can only be formed by wind accretion, then their production could
be far less efficient than for CEMP stars.

\subsection{Could mass transfer rates be affected C- vs. N-richness?}

Our understanding of mass loss in AGB stars is seriously incomplete. However,
one variable that may be important is the dust-to-gas ratio, with mass-loss
rates increasing for increasing amounts of dust. Carbon is a major contributor
to dust in AGB stars, and the dust-to-gas ratio has been parameterized by the
C-richness \citep[e.g.,][]{arndt:97}. Therefore, since HBB stars produce less
C-rich AGB envelopes, their mass loss may be lower and therefore the number of
binary companions that are enriched enough to be included in our sample may be
small. However, a lower mass-loss rate, given enough time, could still
conceivably add up to the same amount of enrichment. Additionally,
\citet{arndt:97} found only a small dependency on C-enhancement.

\subsection{Does hot bottom-burning occur in intermediate mass AGB stars?}

Another explanation for the absence of NEMP stars in our sample could be the
inability of more massive AGB stars (maybe only at very low metallicity) to
produce N via HBB. The efficiency of envelope convection, parameterized by the
mixing length parameter, determines the efficiency of HBB in stellar evolution
models. A larger convective efficiency would imply even more efficient HBB,
but could decrease the dredge-up efficiency, in particular if significant
convective overshoot is present. This would limit the ability of massive AGB
stars to produce primary N. A smaller convective efficiency would lead to less
efficient HBB, and a smaller mass range for efficient N production, lowering
the expected number of NEMP stars. However, we have not performed calculations
with a range of convective efficiencies, and therefore we cannot quantify
these possible effects. Stellar evolution modeling of HBB in massive AGB stars
also predicts the production of Li through the \citet{cameron:71} mechanism.
This property of HBB is observationally confirmed by the discovery that almost
all of the luminous AGB stars with ${\rm C/O} < 1$ in the Magellanic Clouds
are lihtium rich \citep[e.g.][]{smith:95}.

\subsection{Could extra-mixing be affecting our results?}

The main goal of our study was to establish whether there existed a possible
observational selection bias against the discovery of NEMP stars. However, we
also noted that previous literature data, as well as our new objects have C/N
ratios that are too low compared to standard low-mass AGB models without HBB.
Our 2\,{\solmass} and 3\,{\solmass} model sequences produce large amounts of
C, but no N.  The most likely solution to this discrepancy would be an
additional mixing process, such as cool-bottom processing, which has been
studied in great detail in the context of abundance anomalies in globular
cluster stars \citep{wasserburg:95,denissenkov:03b}. Such an extra-mixing
process below the bottom of the convective envelope could also operate in
low-mass AGB stars, leading to mild CN cycling that produces some N, but not
as complete as in HBB \citep{nollett:03}. We believe that the present
observational data support such a scenario. We plan to test this idea
quantitatively in the future with stellar evolution models that include this process.

\subsection{Final Thoughts}

It is probable that the formation of C-rich stars involves a delicate balancing act
in which the N-rich stars are disadvantaged. The most likely possibilities are
the binary mass ratios in the early Universe or the process of mass transfer
for large mass-ratio systems. With ever larger samples of metal-poor stars with
sufficient spectroscopic data becoming available, we expect to continue studies
of the relative fractions of CEMP vs. NEMP stars. To confirm that the majority
of the CEMP stars in our sample were indeed polluted by AGB progenitors, we and
other authors are obtaining high-resolution spectra for many of these stars to
look for s-process enhancements, and to check for detectable radial-velocity
variations. In addition, other elements, such as Li, Na, F, and the isotopes of
Mg, which are affected by nucleosynthesis in AGB stars, will be examined in due
course.

\acknowledgements

T.C.B. acknowledges partial funding for this work from grants AST 00-98508,
AST 00-98549, and AST 04-06784, as well as from grant PHY 02-16783: Physics
Frontiers Center/Joint Institute for Nuclear Astrophysics (JINA), all from the
U.S. National Science Foundation.This publication makes use of data products
from the Two Micron All Sky Survey, which is a joint project of the University
of Massachusetts and the Infrared Processing and Analysis Center/California
Institute of Technology, funded by the National Aeronautics and Space
Administration and the National Science Foundation. This work was carried
out in part under the auspices of the National Nuclear Security 
Administration of the U.S.\ Department of Energy at Los Alamos
National Laboratory under Contract No. DE-AC52-06NA25396, and funded by
the LDRD program
(20060357ER). N.C. acknowledges support from
Deutsche Forschungsgemeinschaft under grants Ch~214/3 and Re~353/44.

\end{document}